\documentclass[epj,final]{svjour}
\usepackage{graphicx}
\usepackage{epsfig}
\usepackage{amsmath,amssymb}

\usepackage{color}

\begin{document}

\authorrunning{F. Fehrer et al}
\titlerunning{Hindered Coulomb explosion of embedded Na clusters ...}
\title{Hindered Coulomb explosion of embedded Na clusters --
  stopping, shape dynamics and energy transport}
\author{F. Fehrer\inst{1} \and P.M. Dinh\inst{2} \and M. B\"ar\inst{1} \and P.-G. Reinhard\inst{2} 
\and E. Suraud\inst{1}} 
\institute{ Laboratoire de Physique Th\'eorique, Universit\'e P. Sabatier,
118 Rte de Narbonne, F-31062 Toulouse cedex, France \and 
Institut f{\"u}r Theoretische Physik, Universit{\"a}t Erlangen,
Staudtstrasse 7, D-91058 Erlangen, Germany}
\date{\today / Received: date / Revised version: date}
\abstract{ 
We investigate the dynamical evolution of a Na$_8$ cluster embedded in
Ar matrices of various sizes from $N=30$ to 1048. 
The system is excited by an intense short laser pulse leading to high
ionization stages.
We analyze the subsequent highly non-linear motion of cluster and
Ar environment in terms of trajectories, shapes, and energy flow.
The most prominent effects are~: 
temporary stabilization of high charge states for several ps, 
sudden stopping of the Coulomb explosion of the embedded Na$_8$
  clusters associated with an extremely fast energy transfer to the Ar
  matrix,
fast distribution of energy throughout the Ar layers by a sound  wave.
Other ionic-atomic transfer and relaxation processes proceed at slower 
scale of few ps. The electron cloud is almost thermally decoupled 
from ions and thermalizes far beyond the ps scale.
} 

\maketitle
\section{Introduction}

Metal clusters have many interesting properties which made them a much
studied species in the past decades. Their valence electron cloud
provides a finite fermion system with remarkable properties as
electronic shell effects and strong optical absorption in a narrow
frequency band, the Mie surface-plasmon resonance, for early reviews
see \cite{Hee93,Bra93,Kre93}. The resonance depends sensitively on the
geometry of the cluster and thus provides an ideal handle for
analyzing and for controlling cluster dynamics. This has been much
exploited for free clusters in all dynamical regimes, for a recent
overview see \cite{Rei03a}. 

The dynamical scenarios become more involved 
when clusters stay in contact with
a substrate, either embedded inside or deposited on a surface. The
variety of combinations grows huge. This is generally viewed as an
advantage because it allows to tailor desired properties, e.g., for
designing bio-markers exploiting the prominent optical properties of
clusters \cite{May01,Dub02}. Corresponding to the huge manifold of
possibilities, there is a huge body of publications. But comparatively
little has been done yet in the regime of highly non-linear dynamics
induced by intense laser fields. 

It is the aim of the present contribution to explore, from a
theoretical perspective, that non-linear regime. As test case, we take Na
clusters because these are conceptually the cleanest metal clusters,
least plagued by interference with core electrons and also the
simplest for a theoretical description. The clusters are embedded in
an Ar substrate. This is an insulator and so we simulate a typical
scenario for a chromophore in an inert environment. Moreover, Ar is a
very soft material having a weak interface energy. It exerts only a
very small perturbation on the cluster which, in turn, maintains
basically its structure and optical properties. That feature had been
exploited in the past for studying optical properties of neutral
clusters which are otherwise hard to handle as free clusters, see e.g.
\cite{Har93,Die02}. The weak-perturbative situation will change if we
consider violent dynamics where cluster electrons and ions are driven
to heftier encounters with the surrounding substrate atoms. 
The actual test case will be Na$_8$ embedded in a sufficiently large
cavity of the Ar$_{N}$ system. We will consider various Ar system sizes
$N$. This serves a double purpose~: large $N$ serve as an approximation
to bulk Ar and comparison with smaller $N$ allows to work out typical
effects of mixed clusters, which are as such an interesting species.
We excite the embedded cluster by intense and short laser pulses to
high charge states (Na$_8^{Q+}$ up to $Q=4$)
and follow the subsequent dynamical
evolution over up to 8 ps. We will analyze the results in terms of
detailed views, energy transfers, evolution of global shape, and
relaxation times.

Clusters in contact with a substrate are much more complex systems
than free clusters. The expense for a fully detailed theoretical
description grows huge. Such
calculations are undertaken where details count, mostly for studying
structure of small compounds (see e.g \cite{Hak96b,Mos02a,Sie04a} for
deposited and embedded metal clusters in various material
combinations). But these subtle models are hardly applicable to a
truly dynamical situation and the enormous effort is not really needed
for the sort of exploratory studies which we have in mind here. The
large difference between metal clusters (reactive) and rare gas
(inert) suggests a hierarchical model where the environment is handled
at a lower level of description.  Such approaches are widely used in
theoretical chemistry under the label
Quantum-Mechanical/Molecular-Mechanical model (QM/MM), see
e.g. \cite{Gre99a}, and in surface physics, see e.g., \cite{Nas01a}. A
model in that spirit for Na clusters in contact with Ar was proposed
in \cite{Ger04b} and applied to structure, optical response and
moderately non-linear dynamics of embedded clusters in
\cite{Feh05a,Feh05b}. Here we are going to carry forth the studies in
the regime of strong laser excitations inducing high charge states and
large rearrangements of cluster and matrix structure.

\section{Formal framework}

\subsection{Short summary of the model}

In order to allow for a sufficiently large Ar matrix, we use a
hierarchical approach. We sketch it here briefly and refer to
\cite{Ger04b,Feh05c} for a detailed layout.

The Na cluster is treated in full microscopic
detail using quantum-mechanical single-particle wavefunctions
$\{\varphi_n({\bf r},t),n=1,\ldots,N_{\rm el}\}$
for
the valence electrons. These are coupled non-adiabatically to classical
molecular dynamics (MD) for the Na$^+$ ions
which are described by their positions
$\{{\bf R}_I,I=1,\ldots,N_{\rm Na}\}$.
In the present paper, we have $N_{\rm el}=N_{\rm Na}=8$.
The electronic wavefunctions are propagated by time-dependent
local-density approximation (TDLDA).
The electron-ion interaction in the cluster is described by soft,
local pseudo-potentials. This TDLDA-MD has been widely validated for linear
and non-linear dynamics of free metal clusters \cite{Rei03a,Cal00}.

Two classical degrees-of-freedom are associated with each Ar atom~:
center-of-mass
$\{{\bf{R}}_a,a=1,\ldots,N\}$, 
and electrical dipole moment which is parameterized as
$\{{\bf R}'_a={\bf R}_a+{\bf d}_a,a=1,\ldots,N\}$.
Note that the Ar dipole is practically handled by two constituents
with opposite charge, positive Ar core (at ${\bf R}_a$) and negative
Ar valence cloud (at ${\bf R}'_a$).
With the atomic dipoles, we explicitely treat the dynamical
polarizability of the atoms through polarization potentials
\cite{Dic58}. Smooth, Gaussian charge distributions are used for Ar
ionic cores and electron clouds in order to regularize the singular
dipole interaction.
We associate a Gaussian charge distribution to
both constituents having a width of the order of the 3p shell in Ar,
similar as was done in \cite{Dup96}. The Coulomb field of the
(softened) Ar dipoles provides the polarization potentials which
are the dominant agents at long range.
The Na$^+$ ions of the metal cluster have net
charge $q_{\rm Na}=+1$ and interact with the Ar dipoles predominantly
by the monopole moment. The small dipole polarizability of the Na$^+$
core is neglected.  The cluster electrons do also couple naturally to
the Coulomb field generated by the atoms.
%
%

The model is fully specified by giving the total energy of the system.
It is composed as
\begin{equation}
\label{eq:etotal}
  E_{\rm total}
  =
  E_{\rm Na cluster}
  +
  E_{\rm Ar}
  +
  E_{\rm coupl}
  +
  E_{\rm VdW}
  \quad.
\end{equation}
The energy of the Na cluster $E_{\rm Na cluster}$ consists of
TDLDA (with a self-interaction correction, see Sec.~\ref{sec:lim}) for
the electrons, MD for ions, and a coupling of 
both by soft, local pseudo-potentials; that standard treatment is
well documented at many places, e.g. \cite{Rei03a,Cal00}.
The Ar system and its coupling to the clusters are described by
\begin{eqnarray}
\label{eq:ar_mat}
  E_{\rm Ar}
  &=&
  \sum_a \frac{{\bf P}_a^2}{2M_{\rm Ar}} 
  +
  \sum_a \frac{{{\bf P}'_{a}}^2}{2m_{\rm Ar}}
  +
  \frac{1}{2} k_{\rm Ar}\left({\bf R}'_{a}-{\bf R}_{a}\right)^2
\nonumber\\
  &&  
  +
  \sum_{a<a'}
  \Big[
    \int\!\! d{\bf r}\rho_{{\rm Ar},a}({\bf r})
    V^{\rm(pol)}_{{\rm Ar},a'}({\bf r})
\nonumber\\
  &&\qquad\qquad
    +
    V^{\rm(core)}_{\rm ArAr}({\bf R}_a\!-\!{\bf R}_{a'})
  \Big]
  \;,
\end{eqnarray}
\begin{eqnarray}
\label{eq:ecoupl}
  E_{\rm coupl}
  &=&
  \sum_{I,a}\left[
    V^{\rm(pol)}_{{\rm Ar},a}({\bf R}_{I})
    +
    V'_{\rm NaAr}({\bf R}_I - {\bf R}_a)
  \right]
\nonumber\\
  &+&
  \int\!\! d{\bf r}\rho_{\rm el}({\bf r})\sum_a \left[
    V^{\rm(pol)}_{{\rm Ar},a}({\bf r})
    +
    W_{\rm elAr}(|{\bf r}\!-\!{\bf R}_a|)
  \right]
  \,,
\end{eqnarray}
\begin{eqnarray}
  V^{\rm(pol)}_{{\rm Ar},a}({\bf r})
  &=&
  e^2{q_{\rm Ar}^{\mbox{}}}
  \Big[
   \frac{\mbox{erf}\left(|{\bf r}\!-\!{\bf R}^{\mbox{}}_a|
          /\sigma_{\rm Ar}^{\mbox{}}\right)}
        {|{\bf r}\!-\!{\bf R}^{\mbox{}}_a|}
\nonumber\\
  &&\qquad
   -
   \frac{\mbox{erf}\left(|{\bf r}\!-\!{\bf R}'_a|/\sigma_{\rm Ar}^{\mbox{}}\right)}
        {|{\bf r}\!-\!{\bf R}'_a|}
  \Big]
  \;,
\label{eq:Arpolpot}
\\
  W_{\rm elAr}(r)
  &=&
  e^2\frac{A_{\rm el}}{1+e^{\beta_{\rm el}(r - r_{\rm el})}}
  \;,
\label{eq:VArel}\\
  V_{\rm ArAr}^{\rm (core)}(R)
  &=& 
  e^2 A_{\rm Ar}\Bigg[
  \left( \frac{R_{\rm Ar}}{R}\right)^{12}
 -\left( \frac{R_{\rm Ar}}{R}\right)^{6}
  \!\Bigg]
  \;,
\label{eq:VArAr}
\\
  V'_{\rm ArNa}(R)
  &=&
  e^2\Bigg[
  A_{\rm Na} \frac{e^{-\beta_{\rm Na} R}}{R}
\nonumber\\
  &&\qquad
  -
  \frac{2}{1\!+\!e^{\alpha_{\rm Na}/R}}
  \left(\frac{C_{\rm Na,6}}{R^6}\!+\!\frac{C_{\rm Na,8}}{R^8}\right)
  \Bigg]
  \;,
\label{eq:VpArNa}
\\
  &&
  4\pi \rho_{{\rm Ar},a}
  =
  \Delta V^{\rm(pol)}_{{\rm Ar},a}
  \;,
\end{eqnarray}
\begin{eqnarray}
  E_{\rm VdW}
  &=&  
  \frac{e^2}{2} \sum_a \alpha_a
  \Big[
    \frac{
       \left(\int\!{d{\bf r} {\bf f}_a({\bf r}) \rho_{\rm el}({\bf r})}\right)^2
         }{N_{\rm el}}
\nonumber\\
   &&
     -\int\!\!{d{\bf r} {\bf f}_a({\bf r})^2 \rho_{\rm el}({\bf r})}
  \Big]
  \,,
\label{eq:EvdW}
\\
  &&
  {\bf f}_a({\bf r})
  =
  \nabla\frac{\mbox{erf}\left(|{\bf r}\!-\!{\bf R}^{\mbox{}}_a|
          /\sigma_{\rm Ar}^{\mbox{}}\right)}
        {|{\bf r}\!-\!{\bf R}^{\mbox{}}_a|}
  \;,
\label{eq:effdip}
\\
  &&
  \mbox{erf}(r)
  = 
  \frac{2}{\sqrt{\pi}}\int_0^r dx\,e^{-x^2}
  \quad.
\end{eqnarray}
The calibration of the various contributions is taken from independent
sources, except eventually for a final fine tuning to the NaAr dimer
modifying only the term $W_{\rm elAr}$ of Eq.(\ref{eq:VArel}). The
parameters are 
summarized in  table \ref{tab:params}. The third column of the table
indicates the source for the parameters. In the following, we report
briefly the motivations for the choices.

\begin{table*}
\begin{center}
\begin{tabular}{|l|l|l|}
\hline
 \rule[-8pt]{0pt}{22pt}
 $V^{\rm(pol)}_{{\rm Ar},a}$
&
 $q_{\rm Ar}
  =
  \frac{\alpha_{\rm Ar}m_{\rm el}\omega_0^2}{e^2}$
 \;,\;
 $k_{\rm Ar}
  =
  \frac{e^2q_{\rm Ar}^2}{\alpha_{\rm Ar}}$
 \;,\;
 $m_{\rm Ar}=q_{\rm Ar}m_{\rm el}$\;,
&
 $\alpha_{\rm Ar}$=11.08$\,{{\rm a}_0}^3$
\\
 \rule[-12pt]{0pt}{22pt}
 &
 $\sigma_{\rm RG}
  =
  \left(\alpha_{\rm Ar}\frac{4\pi}{3(2\pi)^{3/2}}  \right)^{1/3}$
&
 \raisebox{12pt}{$\omega_0 = 1.755\,{\rm Ry}$}
\\
\hline
 \rule[-6pt]{0pt}{18pt}
 $W_{\rm elAr}$
&
 $A_{\rm el}$=0.47  
 \;,\;
 $\beta_{\rm el}$=1.6941\,/a$_0$  
 \;,\;
 $r_{\rm el}$=2.2 a$_0$ 
&
 fit to NaAr dimer
\\ 
\hline
 \rule[-8pt]{0pt}{22pt}
$V^{\rm(core)}_{\rm ArAr}$
&
 $A_{\rm Ar}$=$1.367 \times 10^{-3}$ Ry
 \;,\;
 $R_{\rm Ar}$=6.501 a$_0$ 
&
fit to bulk Ar
\\
\hline
 \rule[-6pt]{0pt}{18pt}
 $V'_{\rm ArNa}$
&
 $\beta_{\rm Na}$= 1.7624 a$_0^{-1}$
 \;,\;
 $\alpha_{\rm Na}$= 1.815 a$_0$
 \;,\;
 $A_{\rm Na}$= 334.85, 
&
\\
 \rule[-6pt]{0pt}{18pt}
&
 $C_{\rm{Na},6}$= 52.5 ${{\rm a}_0}^6$
 \;,\;
 $C_{\rm Na,8}$= 1383 ${{\rm a}_0}^8$
&
after \cite{Rez95}
\\
\hline
\end{tabular}
\end{center}
\caption{\label{tab:params}
Parameters for the various model potentials. See text for detail.
}
\end{table*}

Most important are the polarization potentials defined in
Eq.(\ref{eq:Arpolpot}). They are described by the
model of a valence electron cloud oscillating against the rare gas core
ion. Their parameters are~: 
$q_{\rm Ar}$ the effective charge of valence cloud, 
$m_{\rm Ar}=q_{\rm Ar}m_{\rm el}$ the effective mass of valence cloud, 
$k_{\rm Ar}$ the restoring force for dipoles, and 
$\sigma_{\rm Ar}$ the width of the core and valence clouds.
The $q_{\rm Ar}$ and $k_{\rm Ar}$ are adjusted to reproduce the dynamical
polarizability $\alpha_D(\omega)$ of the Ar atom at low frequencies,
i.e. we choose to reproduce the static limit 
$\alpha_D(\omega\!=\!0)$ and its second derivative
$\alpha''_D(\omega\!=\!0)$.
The width $\sigma_{\rm Ar}$ is determined consistently such that the
restoring force from the folded Coulomb force (for small
displacements) reproduces the spring constant $k_{\rm Ar}$.

The short range repulsion is provided by the various core potentials.
For the Ar-Ar core interaction, Eq.(\ref{eq:VArAr}), we employ a
Lennard-Jones type 
potential with parameters such that binding properties of bulk Ar are
reproduced.  The Na-Ar core potential  Eq.(\ref{eq:VpArNa})
is chosen according to
\cite{Rez95}. Note that the Na-Ar potential from \cite{Rez95} does
also contain a long range part $\propto\alpha_{\rm Ar}$ which accounts
for the dipole polarization-potential. Since we describe that long range part
explicitely, we have to omit it here. We thus choose the form 
as given in $V'_{\rm ArNa}$.

The pseudo-potential $W_{\rm elAr}$,  Eq.(\ref{eq:VArel}),
for the electron-Ar core repulsion
has been modeled according to the proposal of \cite{Dup96}. Its
parameters determine sensitively the binding properties of Na to the
Ar atoms.  We use them as a means for a final fine-tuning of the
model. The benchmark for adjustment is provided by the Na-Ar
dimer. The data (dimer binding energy, 
bond length, excitation energy) are taken from \cite{Gro98,Rho02a}.
The adjustment shows some freedom in the choice of $A_{\rm el}$. We
exploit that to produce the softest reasonable core potential.

The Van-der-Waals energy, $E_{\rm VdW}$,  Eq.(\ref{eq:EvdW})
stems from a correlation of the
dipole excitation in the Ar atom coupled with a dipole excitation in
the cluster. We exploit that the plasmon frequency $\omega_{\rm Mie}$
is far below the excitations in the Ar atom.
This simplifies the term to the variance of the dipole operator in
the cluster, using again the regularized dipole operator ${\bf f}_a$
corresponding to the smoothened Ar charge distributions. 

\subsection{Validity and limitations}
\label{sec:lim}

A few words are in order about the range of validity
for that hierarchical model. The structure of the coupled systems is
well described by construction, see the predecessor in \cite{Dup96}
and the extensive testing in \cite{Ger04b}. The same holds for optical
response (see \cite{Feh05a,Ger04a}). However, one has to remain aware
that the model has limitations with respect to allowed frequencies and
amplitudes. The dynamical response of the Ar atoms is tuned to
frequencies safely below the Ar resonance peak, i.e. safely below 15
eV. This is well fulfilled in the present calculations where the
highest relevant frequency is the cluster's Mie plasmon at around 3
eV.  The amplitudes of the dipole oscillations in Ar should also stay below
the threshold where free electrons are released from the Ar atom into
the matrix. That could become critical for the violent processes
studied here.  We checked the threshold for electron emission by fully
quantum mechanical calculations of laser excitation on Ar atoms and we
find a critical field strength of order of 1.4 eV/a$_0$. That value
holds for constant fields. The case is more forgiving if the critical
field strength is exceeded only for a short time.  Anyway, we protocol
during our calculation the actual field strength at each Ar site.
We are on  the safe side for the cases with charge $Q=3$ and lower.
The dynamics with charge $Q=4$ shows occasionally field peaks
above the limit, however for very short times. 

The quantum mechanical treatment of the Na cluster involves two
approximations which may limit the quantitative value of the results.
The electron cloud is treated with axially averaged potentials and 
the TDLDA is augmented by a self-interaction correction (SIC) at 
the level of average-density SIC (ADSIC) \cite{Leg02,And02b}. Both
approximations enhance the barriers for fragmentation of the cluster,
or pieces thereof. The axial approximation is needed to enable the
large scale calculations performed here. The SIC is compulsory for an
appropriate description of ionization. We pay the price that our
calculations provide rather a qualitative picture. The effects shown
are certainly relevant. The thresholds  concerning charge state
and laser field strength are probably not too precise.
However we also dispose of a full three-dimensional treatment for the
valence electrons which, moreover, can handle various levels of SIC
and even exact exchange. We used that for occasional counterchecks and
we never found significant (i.e. qualitative) deviations from the
approximate treatment.

\subsection{Initialization and propagation}

The test cases in the following studies are Na$_8$ as free cluster
and embedded in Ar$_{N}$ where $N=30$, $164$, $434$, and $1048$. The
Ar structures contain a cavity equivalent to 13 Ar places.  This was
found to provide sufficient space for convenient embedding of Na$_8$.
The latter consists of two rings carrying 4 Na ions each. The
rings have the same diameter and are twisted by  45$^\circ$ 
relative to each other in order to minimize Coulomb energy. That
structure is the same in all cases. The effect of the Ar surroundings
on the structure is very small. For more details on structure and
basic optical properties see \cite{Feh05a}.
All calculations start from fully relaxed stationary ground state
configurations. A laser field is applied as time dependent dipole
field with frequency $\omega_{\rm laser}=2$ eV and a temporal $\sin^2$
envelope with full width at half maximum (FWHM) of 33 fs. The intensity
($\equiv$ field strength) is varied to reach the different final
charge stages $Q=2$, 3, and 4. It has typical values in the range
1--5$\times 10^{12}$W/cm$^2$. Its polarization is along the 
symetry axis ($z$ axis) of the system. 

The numerical treatment uses the standard techniques of
the time-dependent local-density approximation for the cluster
electrons and molecular dynamics for Na$^+$ ions or Ar atoms (TDLDA-MD)
as outlined e.g. in \cite{Rei03a,Cal00}. The wavefunctions and fields
are represented on a grid in coordinate space. The quantum mechanical
time stepping is done  by the time-splitting method and the molecular
dynamics by velocity Verlet. We use absorbing boundary conditions to
simulate correctly electron escape.

\section{Charge stabilization}
\label{sec:stabil}

The laser irradiation leads to strong ionization of the cluster. An
interesting issue is then the Coulomb-stability of the
cluster. This has been much studied in the past under the key word
``appearance size'', the critical size above which clusters of a
certain charge state become ultimatively stable (i.e. appear in a mass
spectrograph), for a review see \cite{Nae97}. It was shown that the
value depends sensitively on the excitation mechanism, with ns laser
pulses being less favorable, and fs laser pulses as well as fast ion
collisions producing significantly lower critical values
\cite{Nta02,Rei00a}. For Na clusters, the smallest Na cluster with
charge state $Q=2$ has been observed for Na$_{10}^{++}$ in experiments
with laser pulses \cite{Sch97a}.
 The case of free Na$_{8}^{++}$, of interest here, 
is at the edge of stability.
Higher charge states $Q\geq 3$, however, lead for sure to immediate
Coulomb explosion of free clusters. The interesting question is then
how the surrounding Ar environment modifies the charge stability. We
will address this question first in terms of asymptotic stability by
energetic estimates and then analyze detailed dynamical processes. We will
see that the Ar matrix manages to prevent immediate Coulomb
fragmentation for surprisingly high charge states. The strong
oscillations of the imprisoned clusters will be analyzed in terms of
global shape parameters.

\subsection{Stability in principle}

TDLDA-MD simulations for free Na$_8^{Q+}$ clusters lead to immediate
Coulomb explosion for charge states $Q\geq 3$. The case of Na$_8^{++}$
is at the fringe of stability where small changes in the conditions
produce large effects in the results. Monte-Carlo cooling still
produces a stable system. The results of dynamical simulations
starting from neutral Na$_8$ and fast ionization depend on the level
of approximation. The calculations with axial approximation for the
electrons show huge shape oscillations with size varying within a
factor of two.  Fully three dimensional treatment (without SIC and
mere TDLDA) starts in the same manner for the first 5 ps and then
``discovers'' the Na$^+$ emission path leading to final Coulomb
fragmentation.   However, the
detailed propagation and the breakup channel depends critically on the
initial configuration. Minimal shifts of the ions can lead into totally
different directions.
\begin{figure}[b]
\centerline{\epsfig{figure=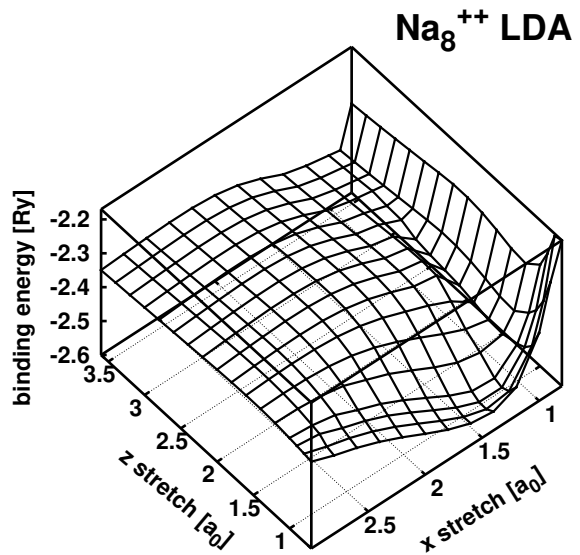,width=8.5cm}}
\caption{\label{fig:na8_LDA}
Born-Oppenheimer surface for free Na$_8^{++}$ in the plane of
collective deformations. The energies are shown
versus the stretching factor in $z$ and $x$ direction.
Three dimensional LDA calculations were used.
}
\end{figure}
It is in any case obvious that the cluster is at least not
instantaneously unstable. This observation is corroborated by
computing (with the 3D code) the Born-Oppenheimer surfaces in the
space of collective deformation (monopole, quadrupole). We find indeed
a local minimum for slightly oblate Na$_8^{++}$ 
surrounded by reaction barriers against fission, as can be seen from
Fig.~\ref{fig:na8_LDA}. The lowest barrier in that space is about 1 eV.
 However, the reaction barrier disappears in 
symmetry breaking directions with single-ion escape.  The Coulomb
excitation by immediate ionization thus produces a metastable
state. It requires some time to develop sufficient asymmetry for
breakup.
The example shows that a case like Na$_8^{++}$ which lies  at the
limits of stability imposes high demands on the theoretical
description.  Nonetheless, one can read the result positively: all
approaches agree in predicting a long initial period of
quasi-stability which is well in the range of experimental
observation, e.g. by pump-and-probe techniques assessing the time
evolution  of cluster shape through the time evolution of the Mie
plasmon resonance \cite{And02,And04,Doe05a}.

\begin{figure}
\centerline{\epsfig{figure=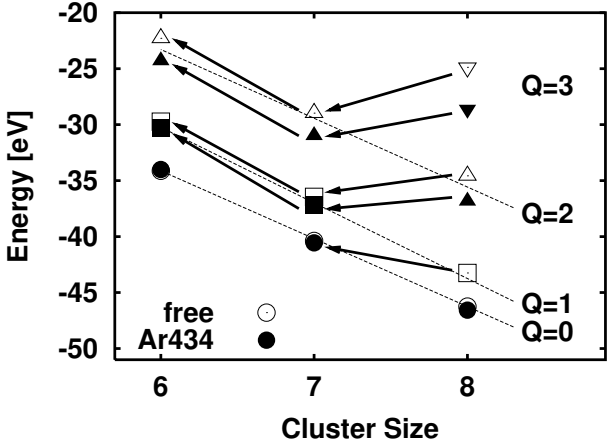,width=\linewidth}}
\caption{\label{fig:energy-trends} 
Binding energies for Na$_{\rm n}^{Q+}$ clusters, free (open symbols) or
embedded in Ar$_{434}$ (filled symbols). 
The different symbols characterize the charge state~:
$Q=0\leftrightarrow$ circles,
$Q=1\leftrightarrow$ squares,
$Q=2\leftrightarrow$ up-triangles,
$Q=3\leftrightarrow$ down-triangles.
The faint dotted lines
connect clusters along same charge states $Q$. 
The arrows show the energetic path for emission of one Na$^+$ ion.
All clusters
have been fully relaxed into their optimal configuration.
}
\end{figure}
Before proceeding to the dynamical simulations, we want
to calibrate
our expectations by looking at the asymptotic stability.  To that end,
we compute the ground-state binding energies of various Na clusters
for several charge states. The results for free and embedded clusters
are shown in Fig.~\ref{fig:energy-trends}.  
The steady down-slope of
the dotted lines shows that the binding energy for fixed $Q$ increases
in almost constant amounts with the number n of Na$^+$ ions. The energy
difference in vertical direction represent the (adiabatic) ionization
potentials which naturally increase with increasing charge states.
The most interesting part is to look at the decay paths of the highly
charged clusters, going along emission of a Na$^+$ ion (arrows).
Down slope means energy gain and thus asymptotic instability. A system
may be locally stable for quite a while (see the above
discussion). However at long times, the system will find its way to the
configuration with lower energy. Free Na$_8^{++}$ is indeed found to
be asymptotically unstable. The energy difference shrinks to a
negligible amount for embedded Na$_8^{++}$. This system will  then be
stable. We see generally that embedding has a stabilizing effect for
charged clusters. For example, the case of Na$_8^{3+}$ which was
clearly explosive for the free cluster is ``downgraded'' to a
situation which is comparable to free  Na$_8^{++}$.
Energy differences alone however are not fully conclusive for the
stability times. These depend also on the phase space of decay
channels and on the reaction barriers. The latter are certainly
larger for embedded clusters, since a single Na$^+$ would have to find
its way through the closely packed Ar medium.

\subsection{Detailed view}



Let us now turn towards the analysis of the dynamical scenario 
following the irradiation of an embedded Na$_8$. 
The initial reaction to laser excitation is exactly the same as we had
already observed previously for the more moderate cases \cite{Feh05b}.
There is a fast direct emission of electrons. The pattern are the same
for free and for embedded clusters. The Ar environment does not impose
any hindrance, even for the largest system in our sample carrying 1048
Ar atoms. The fast ionization is completed at the end of the pulse,
i.e. at 100 fs, and we are left with a cluster in a certain charge
state $Q$ depending on the laser intensity.
The finite net charge of our test system is related to the finite size
of the setup.  In a macroscopically large matrix, the electrons would
be stuck somewhere within the range of their mean free path.  They may
even drift very slowly back towards the now attractive cluster, and
eventually recombine there. We have simulated that by using reflecting
boundary conditions rather than absorbing ones and we find
recombination times of the order of several 10 ps. Thus the present
scenario should provide a pertinent picture for several ps, the time
window studied here.

\begin{figure*}
\centerline{\epsfig{figure=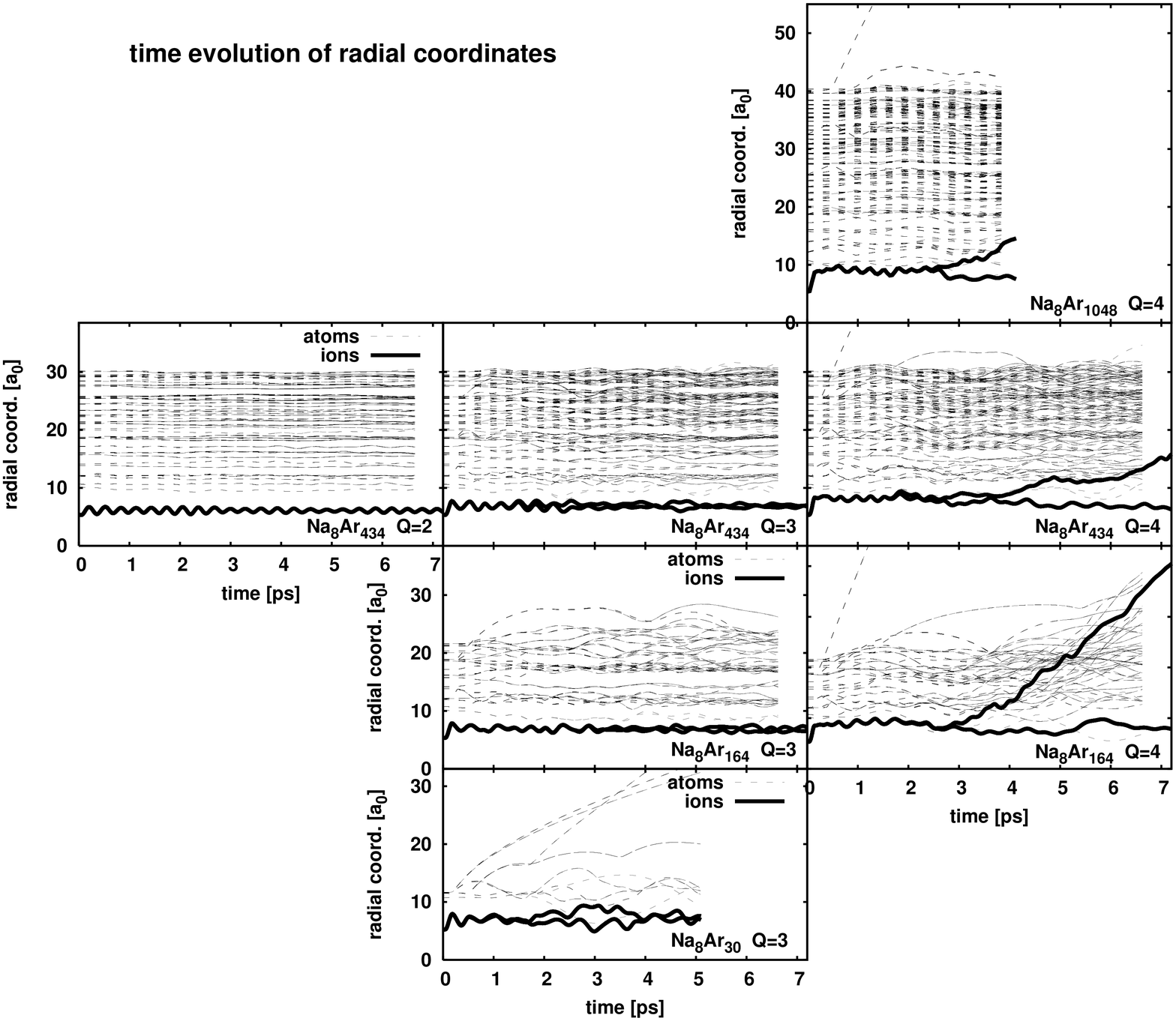,width=18cm}}
\caption{\label{fig:na8_arxx_posrad} 
Time evolution of radial coordinates
$r_n^{\mbox{}}=\sqrt{x_n^2+y_n^2+z_n^2}$
of ions (heavy lines) and atoms (faint lines)
for embedded Na$_8$ after irradiation with
a short laser pulse having $\omega_{\rm laser}=2$ eV and
$\sin^2$ envelope with FWHM = 33 fs. The intensity
is varied to reach the different final charge stages $Q$
as indicated. The plotting of the Ar atoms is stopped a bit earlier
to display more clearly the final status of the Na ions.
}
\end{figure*}
The Coulomb pressure generated by the almost instantaneous ionization
drives the cluster ions apart which, in turn, stir up the Ar atoms.
The dominant effect is expansion. Thus a visualization in terms of
radial coordinates is most appropriate.  This is done in
Fig.~\ref{fig:na8_arxx_posrad} showing the detailed time evolution for
the various charge states and system sizes. 
The cavity is spacious, leaving a large initial separation of the Na
ions from the first shell of Ar atoms. Thus we see for the initial 200
fs a fast expansion/explosion of the cluster almost as in the free
case.
The expansion is stopped instantaneously if the cluster runs against
the repulse Ar core.  The stopping radius increases with the charge
state $Q$. After stopping, the cluster turns over into oscillations
for a while.
The Ar atoms take up the momentum from the stopped ions. That momentum
propagates like a sound wave through the Ar medium.  The perturbation  is
strong enough for $Q=4$ to produce finally a direct emission of Ar
atoms when the wave has reached the outmost shell.  On the way out,
the  wave distributes the energy very quickly over all shells
exciting them to more or less hefty fluctuations. The Ar amplitudes
increase with increasing charge state $Q$ and decreasing system
size. The charge dependence is related to the initial amount of
Coulomb energy $\propto Q^2$. The size dependence comes from the
sharing of energy over the Ar degrees of freedom. More Ar atoms leave
less energy per atom and subsequently a smaller amplitude of motion.
The Ar matrix appears  surprisingly robust for $N=434$ and larger.
Heating suffices for $N=30$, $Q=3$ and $N=164$, $Q=4$ to induce a
steady dissolution of the Ar environment within a few ps. The process
starts from the outer shells (mind that melting starts generally at
the surface \cite{Loe89}) such that the Na cluster remains still
captured by the Ar shells for a long time.  The motion of the
outmost shell shows also interestingly large excursions in other cases
($N=164$ at $Q=3$ and $N=434$ at $Q=4$).  The high net charge of the
total system together with the dipole polarizability of the Ar atoms
generates a long-range attractive potential $\propto r^{-4}$ which
allows for these large fluctuations of still bound atoms.
The stabilization of the cluster appears rather limited for the
highest charge 
state $Q=4$. At about 4 ps, the cluster expansion revives. It proceeds,
however, at a very slow time scale. One could call that a Coulomb
driven diffusion through the medium. Such a process is not visible
within the studied time span (up to 10 ps, shown are 7 ps
for graphical reasons) for charge state $Q=3$.
It cannot be excluded on a longer time span. However, in any case
up to $Q=4$, we see a temporary stabilization inside the cavity which
lasts sufficiently long for an observation.

We can also read off from Fig.~\ref{fig:na8_arxx_posrad} a few time
scales. The initial Coulomb explosion lasts only for about 200 fs.
The distribution of energy over the Ar shells is done within 0.5-0.7
ps. In fact it propagates with the speed of the sound wave which is
about 30 a$_0$/ps for $Q=3$ and even faster for $Q=4$. This is close
to the velocity of sound in pure Ar, as we will see in the next section. 
The temporary stabilization of the cluster lasts about 4 ps for $Q=4$,
a  yet unknown longer time for $Q=3$. 
The thermal dissolution of the Ar system takes several ps starting
from outmost shells.

\begin{figure}
\centerline{\epsfig{figure=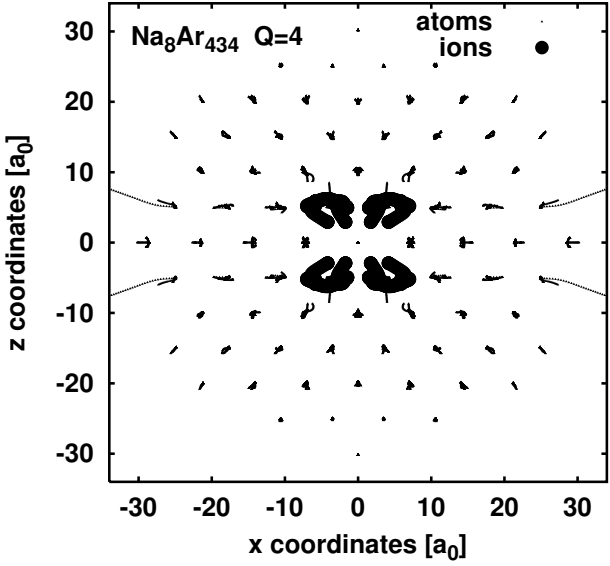,width=8.5cm}}
\caption{\label{fig:na8_arxx_trajectories}
Trajectories in the $x$-$z$-plane for $Q=4$ 
and $N=434$.
}
\end{figure}
The instantaneous stopping, the sound wave and the subsequent direct
emission of Ar atoms are interesting processes which deserve further
inspection. Fig.~\ref{fig:na8_arxx_trajectories} shows a projection
of ionic and atomic trajectories onto the $x$-$z$ plane (where $z$
denotes the symmetry axis, and laser polarization axis).  
The ions start with a radial Coulomb explosion, are
bent over with little energy loss by about 135$^\circ$ and hit almost head
on the next Ar atom. This then triggers the sound wave.  The wave is
not well visible in coordinate space. It is basically a momentum wave
transmitted through small kicks of the atoms. The final kick, however,
releases the atom in the last shell for a long travel. All eight ions
perform the initial looping similarly and eight
Ar atoms are eventually released in that case of $Q=4$ and $N=434$.

\subsection{Global shape}

Complementing information can be gained from characterizing the system
through global shape parameters. The leading quantity is the
overall extension which can be quantified by the root-mean-square
(r.m.s.) radius $r$. Next important is the quadrupole deformation
which  we describe by the dimensionless quadrupole moment $\beta_2$.
Both observables are defined in detail as
\begin{subequations}
\label{eq:shape}
\begin{eqnarray}
  r
  &=&
 \sqrt{\frac{1}{p} \sum_{n=1}^{p} r_n^2}
 \quad,
\\
  \beta_2
  &=&
  \sqrt{\frac{\pi}{5} } \frac{1}{pr^2} \sum_{n=1}^{p} (2z^2-x^2-y^2)
  \quad.
\end{eqnarray}
\end{subequations}
These shape parameters are computed for the Na cluster as well as for
the Ar system. The sum runs up to $p=8$ for the Na cluster and up to
$p=N$, the number of Ar atoms.

\begin{figure}
\centerline{\epsfig{figure=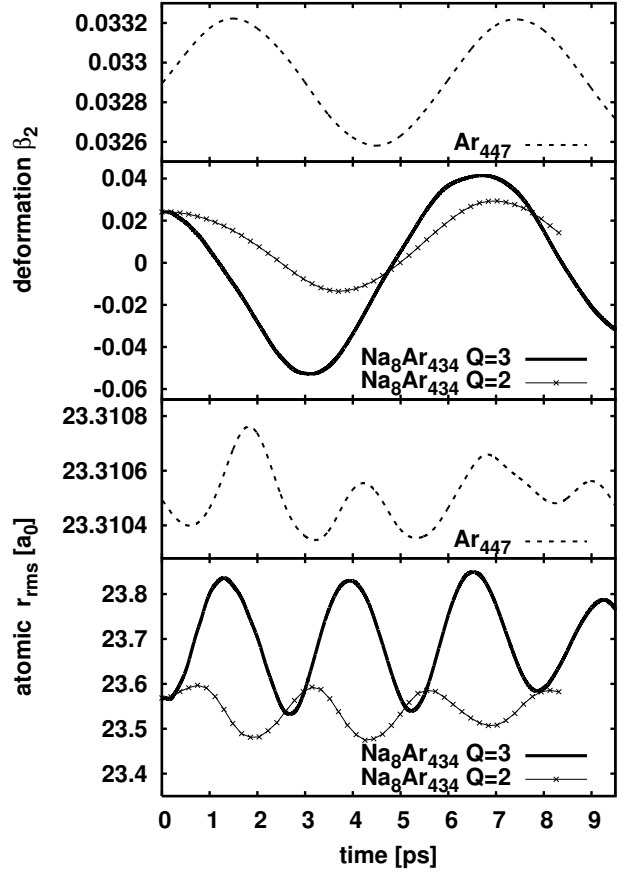,width=8.5cm}}
\caption{\label{fig:ar447_radglobal}
Time evolution of atomic shape parameters 
(\ref{eq:shape}), r.m.s. radius $r$ and deformation
$\beta_2$, for a pure Ar$_{447}$ cluster and
Na$_8$\@Ar$_{434}$ at two charge states.
}
\end{figure}
Fig.~\ref{fig:ar447_radglobal} shows the time evolution of atomic
shape parameters for two charge states in Ar$_{434}$ and compares it
with the corresponding pure Ar cluster Ar$_{447}$ which contains also
the 13 atoms of the cavity. The Ar$_{447}$ was excited by the same
laser pulse as for the case $Q=3$. The laser has a handle on the Ar
atoms through their dynamical dipole polarizability.  The oscillations
have surprisingly much in common. Besides the amplitudes, the pattern
are similar and so are the cycle times, for radial oscillations about
2-3 ps and for quadrupole oscillations about 6-7 ps.  The large charge
at the center of the system seems to change very little on the global
properties. The matrix is also in that sense inert.
%
%
%
The cleanly developed modes allow to estimate the sound velocity in
these large clusters.  The radial compression mode corresponds to the
longitudinal sound mode in bulk material.  Its frequency is
$\omega_{\rm vib}\approx 1.8$ meV.  The momentum of the radial wave is
$q=\pi/R$ where $R=30$ a$_0$ is the cluster radius. The sound velocity
is then estimated as $v_{\rm sound}=\omega_{\rm vib}/q\approx30$
a$_0$/ps which is very close to the propagation speed as observed in
Fig.~\ref{fig:na8_arxx_posrad}.
The quadrupole mode is a surface mode. Its lower frequency of about 0.7
meV yields a velocity of 12 a$_0$/ps corresponding to a
surface sound wave. 

The amplitudes are, of course, different for the various cases (note
the vertical scales in Fig.~\ref{fig:ar447_radglobal}).  The
strong laser field has a very small, although visible, effect on the
pure cluster. The strong reaction for the composite system is due to
the chromophore embedded at the center. This couples strongly to the
laser, becomes ionized, and transfers a large portion of its
excitation to the Ar surroundings. The strength of the initial
excitation (that is, ionization) translates directly into the
amplitude of the indirectly triggered oscillations.

\begin{figure*}
\centerline{\epsfig{figure=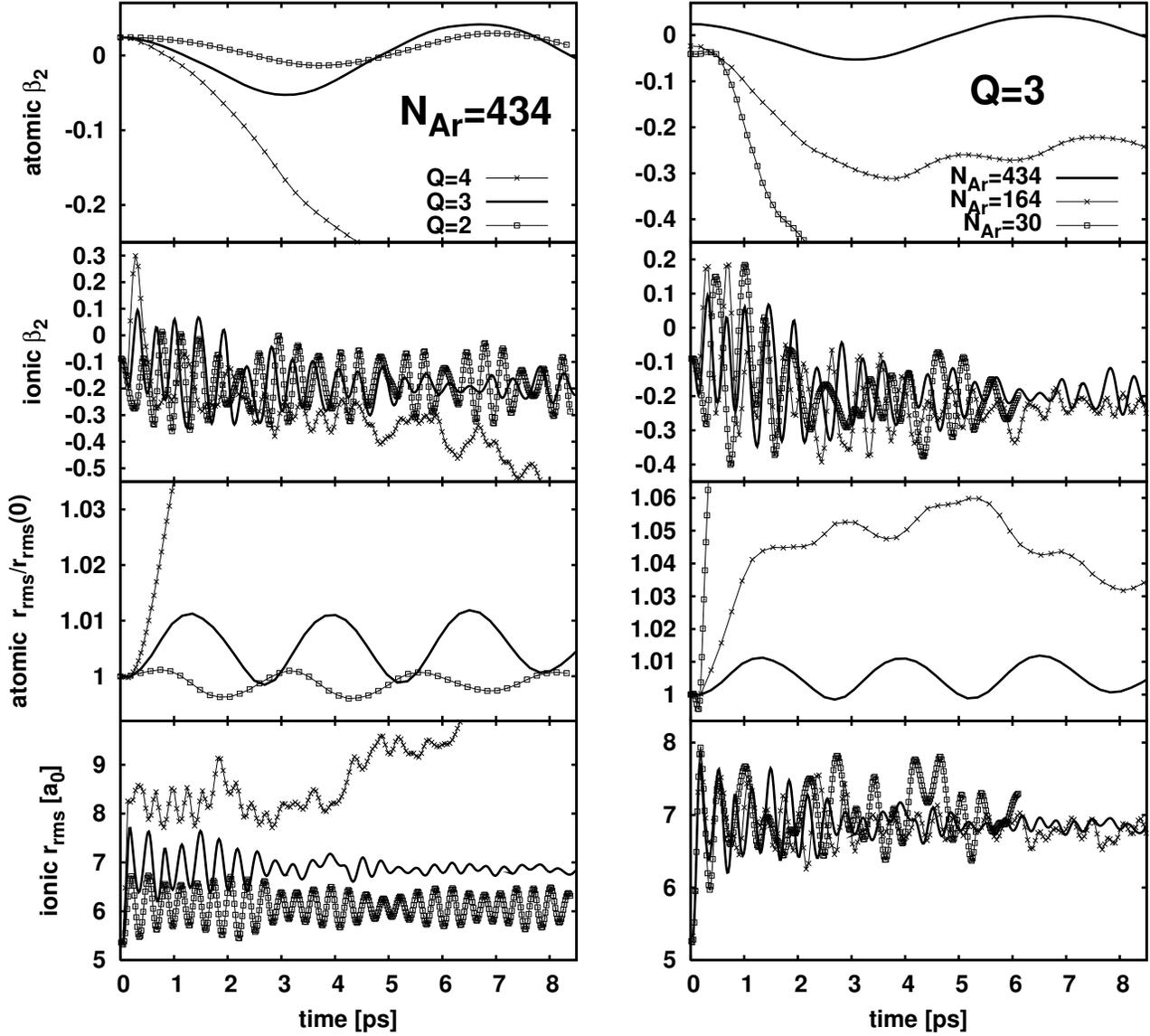,width=17cm}}
\caption{\label{fig:na8_arxx_radglobal}
Time evolution of atomic and ionic shape parameters, Eq.
(\ref{eq:shape}), r.m.s. radius $r$ and deformation
$\beta_2$, for several test cases in comparison. Left panel~:
Variation of charge state $Q$ for matrix size
$N_{\rm Ar}=434$. Right panel~: Variation of matrix size
for charge state $Q=3$.
}
\end{figure*}
Fig.~\ref{fig:na8_arxx_radglobal} summarizes the time evolution of ionic
and atomic shape parameters for a great variety of systems
(left column with $N_{\rm Ar} = 434$ and varying charge $Q$, right 
column with $Q=3$ and varying $N_{\rm Ar}$).  The ionic
radii show again the fast stopping. The stopping radius depends
strongly on the charge state $Q$ (left lower panel)
but seems independent of matrix
size (right lower panel). 
What counts is the first shell of Ar atoms whose radius is the
same in all Ar surroundings \cite{Feh05a}. We also nicely see the
steady oscillations for $Q\leq 3$ and the start of Coulomb diffusion
at 4 ps for $Q=4$. 
Comparing the ionic $r_{\rm rms}$ for $Q=2$ and $Q=3$ (lower left panel),
we see that larger $Q$ leads to stronger damping of the ionic
oscillations because the ions, coming closer to the
Ar atoms, couple more efficiently to the Ar system. The Na$^+$-Ar
relaxation time which is very long for moderate excitations (see
$Q=2$ here and \cite{Feh05b}) shrinks to about 4 ps for $Q=3$.  The
effect of system size on the relaxation (right lower panel) is
small when comparing $N=434$ with 164. The case of $N=30$ is too
chaotic to be conclusive.

The effect on the atomic radii (second lower panel in
Fig.~\ref{fig:na8_arxx_radglobal}) is related to stopping radius
(left) and 
to system size (right). Larger ionization translates to larger
amplitudes up to direct atom emission ($Q=4$). Smaller system size
yields more energy per atom and also increases the amplitudes (right
panel). Particularly impressive is the case of $N=164$ with $Q=3$.
This system shows comparatively huge non-linear fluctuations and is
thus at the onset of melting.

The quadrupole moments (upper two panels) behave analogously.  The Na
ions show a strong initial reaction and then develop in the average a
large oblate deformation (negative $\beta_2$). The damping again 
strongly depends on charge state $Q$ and less on system size.  The final
fragmentation for $Q=4$ goes clearly to the oblate side. Actually the Coulomb
force enhances deformation and we start from the oblate side.  That
means geometrically that the ions depart preferably orthogonal from the
symmetry axis (and thus the laser polarization). 
The trend to the oblate side carries over to the Ar
atoms in unstable situations. Still, the average deformation remains
negligible in the stable cases, showing again the inertness of the Ar
matrix.

\subsection{Bond lengths}

The shape parameters (\ref{eq:shape}) provide a global measure.  A
complementing view inside the system is provided by quantifying the
bonding structure. It is related to the distances between the Ar atoms.
These are fixed in a totally frozen configuration.  The temporal
fluctuations of the distances help to conclude on the stability of a
structure. A convenient measure is the average bond length fluctuation
\begin{equation}
  \delta_{\rm bond} 
  =
  \frac{2}{N(N-1)} \sum_{i<j} \frac{ \left( \langle r_{ij}^2 \rangle
  -\langle r_{ij}\rangle^2 \right)^{1/2}}{\langle r_{ij} \rangle}
\label{eq:bondfluc}
\end{equation}
where $\langle \ldots \rangle$ is a time-average over a large interval.
We use here 1 ps averaging time.
The bond length fluctuation is dimensionless, rescaled by the average bond
distance. Thus it measures selectively the state of order.
A collective deformation (radial expansion, quadrupole deformation)
thus does not contribute significantly.

\begin{figure}
\centerline{\epsfig{figure=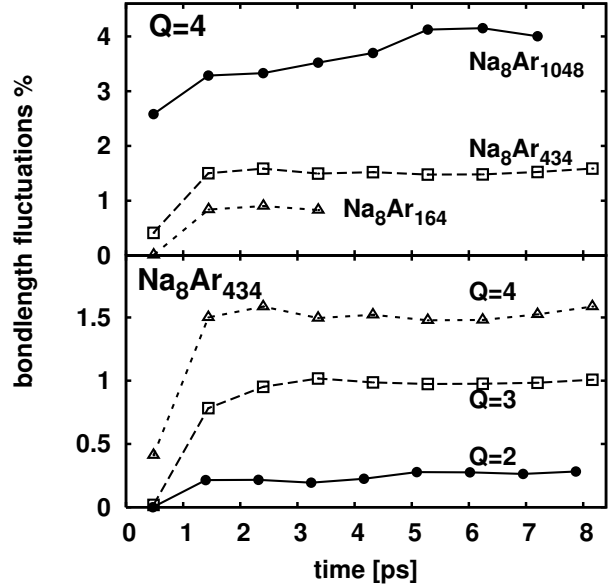,width=8.2cm}}
\caption{\label{fig:na8_arxx_bond}
Time evolution of the bond length fluctuation (\ref{eq:bondfluc}).
Lower panel~: Case of Na$_8$\@Ar$_{434}$ at three different charge states.
Upper panel~: Charge state $Q=4$ for three different matrix sizes.
}
\end{figure}
Fig.~\ref{fig:na8_arxx_bond} shows results for the bond length
fluctuations. They remain surprisingly small in all cases. 
At first glance, this seems to contradict the huge spatial
fluctuations seen in Fig.~\ref{fig:na8_arxx_posrad}. 
Two points are nevertheless to be noted. 
First, the fluctuations start at the outer
shells while the majority of inner Ar atoms remains still better
behaved, and second, the bond length fluctuations measure the
intrinsic order while being insensitive to global motion.
The result indicates that the Ar system stays surprisingly cold in
spite of the sometimes hefty reactions. This conclusion will be
corroborated by an analysis of the internal temperatures in  the
following section.

\section{Energy transfer}

\begin{figure}
\centerline{\epsfig{figure=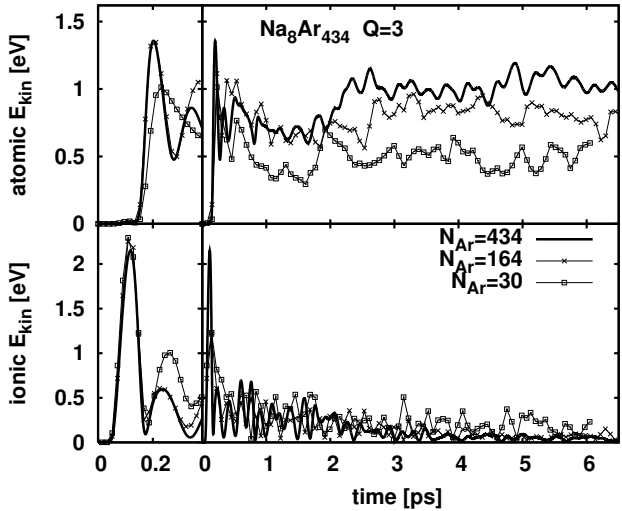,width=8.5cm}}
\caption{\label{fig:na8_arxx_ekinq3}
Time evolution of ionic and atomic kinetic
energies for Na$_8$ in matrices of different size
at charge state $Q=3$. The left panel zooms on the first few 100 fs.
}
\end{figure}
Less visible but equally important observables are the various
energies in the subsystems and the flow between them. In fact, energy
transport and thermalization are most important issues in composite
systems, e.g. for analyzing radiation damage in materials
\cite{Niv00a,Bar02b}. The simplest way to characterize the share of
energies uses the kinetic energies. These are one-body operators and
allow an unambiguous definition for each subsystem, for the cluster
electrons, the ions, and the Ar atoms.
Fig.~\ref{fig:na8_arxx_ekinq3} shows the ionic and atomic kinetic
energies for $Q=3$ and various matrix sizes. The initial phase is
detailed in the left panels. One sees the explosive growth of ionic
energies and the sudden stopping with nearly complete energy exchange
between ions and atoms. About 2/3 of the initial kinetic energy is
kicked over at once to the first Ar shell.  The ions turn to
oscillations and the damping of the oscillations, seen in
Figs.~\ref{fig:na8_arxx_posrad} and \ref{fig:na8_arxx_radglobal},
leads to a steady reduction of ionic kinetic energy. 
The further evolution of the atomic kinetic energy stays below the
first peak. Part of the initial energy gain is invested into
potential energy for spatial rearrangement of atomic positions and
dipoles. Something new happens at about 2 ps, particularly for $N=434$. 
The kinetic energy increases by
about 50\%. Part of that energy comes from ionic relaxation. Another
part has to stem from further spatial rearrangement into an
energetically more favorable configuration. After that, the atomic
energy stays almost constant. All further energy transfer processes
are too slow to be resolved within simulation time.

\begin{figure*}
\centerline{\epsfig{figure=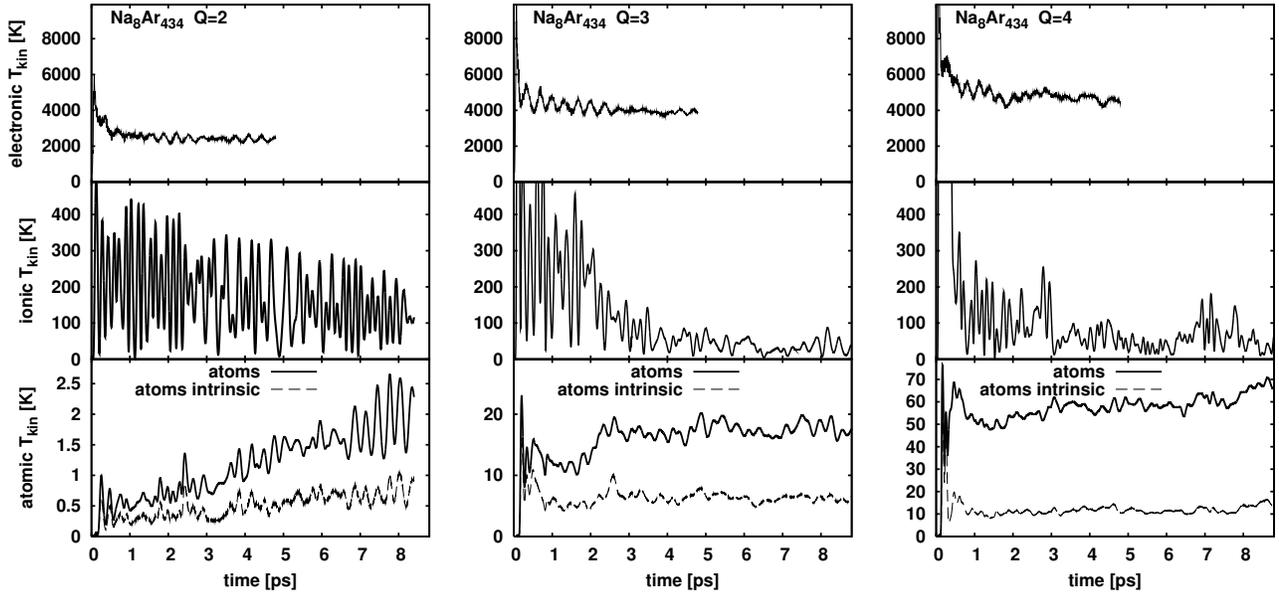,width=17cm}}
\caption{\label{fig:na8_arxx_temp}
Time evolution of electronic, ionic and atomic kinetic
temperatures for Na$_8$\@Ar$_{434}$ at three different charge states.
For the atoms, we show the temperature from total kinetic energy and
from ``intrinsic'' kinetic energy for comparison
(the kinetic energy from collective radial motion is subtracted).
}
\end{figure*}
However total energies are not the best measure for internal excitation.
Energy per particle is better suited. That leads naturally to the
concept of the kinetic temperature
$T_{\rm kin}=2E_{\rm kin}/(3N)$.
The definition is a bit more involved for the electrons.
The kinetic energy as expectation value  of wavefunctions 
contains an unavoidable offset from quantum uncertainty and
from the Pauli principle. We subtract that offset, as well as 
possible collective flow contribution,
to define
an intrinsic electron temperature, following the procedure 
described in \cite{Cal00}.

\begin{figure}
\centerline{\epsfig{figure=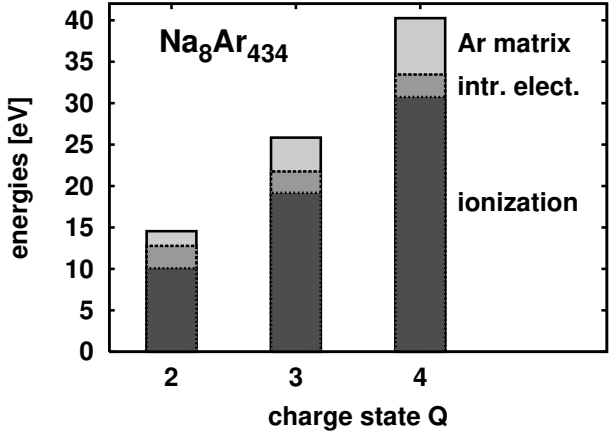,width=8.5cm}}
\caption{\label{fig:balance}
Distribution of the energy initially absorbed from the
laser over the various reaction channels~:
``ionization'' stands for the energy used up to reach the
ionization stage $Q$, ``intr. elect.'' stands for the
intrinsic electronic kinetic energy left in the remaining
electron cloud, ``Ar matrix'' stands for the kinetic and
potential energy stored in the Ar
matrix after relaxation (Na$^+$ ions have then only a very small share).
}
\end{figure}

The various kinetic temperatures are shown for $N=434$ and all three
charge states in Fig.~\ref{fig:na8_arxx_temp}. Note that electrons
can use the same scale for the different $Q$. Their intrinsic
temperature grows with $Q$ but less than linearly. That is due to the
strong cooling through ionization, see the discussion of
Fig.~\ref{fig:balance} later on. In any case, the electronic
temperatures 
are much higher than any other and there is no sign of relaxation towards
the other parts, on the time span explored here. 
This shows that from a  thermal point of view 
electrons and ions/atoms are, to a large extent, still
decoupled. One can conclude that thermal relaxation times 
seem to be far beyond our analysis.
The large electronic temperatures call, in fact, for a description
beyond pure mean field. Electron-electron collisions play a role in
that regime. These could be included by switching to a semi-classical
Vlasov-Uehling-Uhlenbeck description of the electron cloud
\cite{Feh05c,Gig01a}. But long-time semi-classical propagation is
extremely hard to stabilize and, as we have seen above, 
electrons are almost decoupled from ions, thermally speaking. 
Thus we can neglect that effect for the present study, at least on the 
time scales explored here.

The ionic temperatures show somewhat larger variation, still fitting
almost on the same scale. The initial peak (see lower left panel in
Fig.~\ref{fig:na8_arxx_ekinq3}) is not fully shown here. It grows
with a rate between linear and quadratic in $Q$. The further evolution
shows the reverse trend. The kinetic energies shrink the faster the higher
$Q$. That is due to the increasing coupling to the first Ar shell
which accelerates the energy exchange to the  Ar system.

Totally different temperature scales are required for the Ar matrix.
Two effects cooperate here~: The ionic kinetic energy grows with $Q$
and the coupling to the Ar is strongly enhanced with the increasing
ionic stopping radius, see lower left panel of
Fig.~\ref{fig:na8_arxx_radglobal}. The case of $Q=2$ does not show the 
abrupt stopping and thus does not have the large initial jump in
energy. The temperature rather grows steadily in relation to the
equally steady cooling of the ionic cloud. The relaxation time for the
thermalization is of order of 10 ps. That relaxation time drops to 2-3
ps for the larger $Q$ with their more intimate Na$^+$-Ar coupling.
Moreover, most of the energy transfer is concentrated in the stopping
and comparatively little is left for the subsequent energy exchange.
Within the resolution of the results, we can state that Na$^+$-Ar
thermalization is achieved after 4 ps for $Q=3$.  The energy continues to
increase for $Q=4$. That is due to the Coulomb diffusion which
converts steadily potential into kinetic energy.

The results from section \ref{sec:stabil} show a large amount of
collective motion in radial direction and in quadrupole deformation.
That part of motion is still included in the total atomic kinetic
energy. In order to single out the irregular part corresponding to
truly intrinsic excitation, we have computed the collective kinetic
energy contained in the radial oscillations (the slower quadrupole
part contributes very little) and subtracted it to obtain an intrinsic
kinetic energy. The corresponding kinetic temperature is also shown as
the curves labeled ``atoms intrinsic'' in the lower panels
of Fig.~\ref{fig:na8_arxx_temp}.  The
collective part is always the largest portion and even dominates by
far for $Q\geq 3$. 
The minor part is left for intrinsic motion. That
explains the findings from the bond lengths in
Fig.~\ref{fig:na8_arxx_bond} which show very little perturbation of
the internal structure of the Ar system. In spite of the huge amounts of
energy flowing around, that structure persists astonishingly robust.

Finally Fig.~\ref{fig:balance} shows the distribution of the excitation
energy initially induced by the laser.
The total height stands for the excitation energy. Of course, it increases
with increasing intensity and subsequently increasing charge
state. This net investment grows a bit faster than linear.
Earlier experience with free clusters shows that a major cooling
mechanism is electron emission \cite{Cal98c,Rei99a}.  We have
estimated that for the present case by integrating the ionization
potentials (IP) along the ionization curve and by computing the IP as
a function of charge for fixed ionic configuration.  The results in
Fig.~\ref{fig:balance} prove that indeed the by far dominant portion
of energy is carried away with the directly emitted electrons.
That removes typically 80\% of the laser energy.
The next two large contributors are the intrinsic electronic energy
and the final energy of the Ar atoms (which was, to a large
extent, initially ionic kinetic energy). The share depends strongly on
the charge state. Very little is left for the Ar atoms at $Q=2$ while
they gather almost equal share for $Q=4$.  Here we have to keep in
mind that the energy balance in Fig.~\ref{fig:balance} represents
the status at the end of the simulation, i.e. at 6 ps. Final full
thermalization (if reached before dissolution of the system) will give
the kinetic energy of electrons, ions and atoms, with a thermal share
proportional to the number of particles in  each subsystem. 
We have order of 8 electrons and ions but much more atoms.
Thus the Ar system will acquire finally
almost all residual energy which amounts to about 3-6 eV, corresponding to
a heating of about 50-100 K.

\section{Conclusions}

We have analyzed the laser-induced dynamics of a Na$_8$ cluster
embedded in an Ar matrix.  We have used a hierarchical model with a
detailed quantum-mechanical treatment of the clusters electron cloud
and a classical description of the Na$^+$ ions as well as of the
surrounding Ar atoms whereby we include the Ar dipole moments to
account for the dynamical polarizabilities. 
The Ar matrix is approximated by a large Ar cluster covering several
shells of atoms. To distinguish finite size  effects, we
have used different sizes from $N=30$ to 1048 Ar atoms.
The test cases provide several interesting
aspects~: 
They are by construction a model for a cluster embedded in a rare gas
matrix, 
they display also interesting features specific for a finite
composite,
and they are prototypes of a chromophore in an else-wise inert
material.
For the excitation mechanism, we consider irradiation by a short 33 fs
laser pulse with intensity of about $10^{12}$W/cm$^2$.

It was found in earlier studies of structure and mild excitations that
the Ar matrix induces only very small perturbations of the cluster
properties. That remains true for the electronic dynamics in the more
violent regime. However, the now released large amplitude motion of the
cluster ions interferes with the atomic cage. This produces dramatic
differences of the cluster dynamics as compared with the free case. 
The (meta-)stability of charged clusters is much enhanced; even 
a Na$_8^{4+}$ cluster remains bound for about 4 ps, and lower charge
states much longer.
The Coulomb explosion of the highly ionized cluster is abruptly
stopped by the first shell of Ar atoms and most of ionic kinetic
energy is transferred to the Ar shell within a few 10 fs. 
The transferred energy is then spread very quickly by a sound wave
all over the Ar matrix.
Subsequent relaxation and rearrangement processes between ions and
atoms proceed at a time scale of several ps.
The Coulomb instability takes over within that time span for $Q=4$ and
drives a slow Coulomb-driven diffusion of Na$^+$ ions through the Ar
medium.
All processes related to shape evolution within that time scale are in
reach of experimental observation by pump-and-probe analysis using the
cluster surface plasmon as handle.

We have also analyzed the energy balance. The time scales for energy
transfer confirm the findings for the ionic-atomic relaxation as
deduced from shape analysis. The electron cloud adds by far the most
efficient energy loss through direct ionization during laser impact.
And even after large energy outflow with the emitted electrons, a large
part of excitation energy remains kept in internal energy of the
electrons because the thermal (i.e. non-collective) coupling of
electrons to ions and atoms is extremely weak. The small remaining
fraction of excitation energy goes through Coulomb energy (charging),
kinetic energy of Na$^+$ ions (Coulomb explosion), quickly to the Ar
system.  In spite of the small amount of initial laser energy finally
transmitted to the Ar matrix, the Na cluster as chromophore pumps
orders of magnitude more energy into the system as a pure Ar cluster
would be able to extract from the same laser pulse.

\begin{acknowledgement}
This work was supported by the DFG (RE 322/10-1),
the french-german exchange program PROCOPE,  
 the CNRS Programme "Mat\'eriaux" 
  (CPR-ISMIR), Institut Universitaire de France, a
  Bessel-Humboldt prize and a Gay-Lussac prize.
\end{acknowledgement}

\bibliographystyle{epj}  
\bibliography{cluster,add}  

\end{document}